\documentclass[preprint2]{aastex63}

\submitjournal{ApJ}

\shorttitle{Ultrafast wind in Mrk 1044}
\shortauthors{Krongold et al.}
\graphicspath{{./}{figures/}}
\usepackage{graphicx}	
\usepackage[utf8]{inputenc}
\usepackage{wrapfig}
\usepackage{tablefootnote}
\usepackage{amsmath}	
\usepackage{amssymb}	

\newcommand{\pcm}{\mbox{\rm cm$^{-2}$}}	
\newcommand{\xmm}{\mbox{\it XMM--Newton}}
\newcommand{\msun}{\mbox{$M_\odot$}}
\newcommand{\kms}{\mbox{\rm km\,s$^{-1}$}}
\newcommand{\ergsec}{\mbox{\rm erg\,s$^{-1}$}}

\newcommand{\hb}{\mbox{\rm H$\beta$}}
\newcommand{\hi}{\mbox{\rm H\,{\sc i}}}
\newcommand{\civ}{\mbox{\rm C\,{\sc iv}}}
\newcommand{\cvi}{\mbox{\rm C\,{\sc vi}}}
\newcommand{\feii}{\mbox{\rm Fe\,{\sc ii}}}
\newcommand{\fexix}{\mbox{\rm Fe\,{\sc xix}}}
\newcommand{\fexii}{\mbox{\rm Fe\,{\sc xii}}}

\newcommand{\fexx}{\mbox{\rm Fe\,{\sc xx}}}

\newcommand{\fexxii}{\mbox{\rm Fe\,{\sc xxii}}}

\newcommand{\fexxiv}{\mbox{\rm Fe\,{\sc xxiv}}}
\newcommand{\fexxv}{\mbox{\rm Fe\,{\sc xxv}}}
\newcommand{\fexxvi}{\mbox{\rm Fe\,{\sc xxvi}}}
\newcommand{\nixxii}{\mbox{\rm Ni\,{\sc xxii}}}
\newcommand{\nixxiii}{\mbox{\rm Ni\,{\sc xxiii}}}
\newcommand{\nixxiv}{\mbox{\rm Ni\,{\sc xxiv}}}
\newcommand{\naxi}{\mbox{\rm Na\,{\sc xi}}}

\newcommand{\niti}{\mbox{\rm N\,{\sc i}}}
\newcommand{\nii}{\mbox{\rm N\,{\sc ii}}}
\newcommand{\niii}{\mbox{\rm N\,{\sc iii}}}
\newcommand{\niv}{\mbox{\rm N\,{\sc iv}}}
\newcommand{\nv}{\mbox{\rm N\,{\sc v}}}
\newcommand{\nvi}{\mbox{\rm N\,{\sc vi}}}
\newcommand{\nvii}{\mbox{\rm N\,{\sc vii}}}
\newcommand{\nex}{\mbox{\rm Ne\,{\sc x}}}
\newcommand{\oi}{\mbox{\rm O\,{\sc i}}}
\newcommand{\oiv}{\mbox{\rm O\,{\sc iv}}}
\newcommand{\ovi}{\mbox{\rm O\,{\sc vi}}}
\newcommand{\ovii}{\mbox{\rm O\,{\sc vii}}}
\newcommand{\oviii}{\mbox{\rm O\,{\sc viii}}}

\begin{document}

\title{Detection of a Multi-Phase Ultra-Fast Wind in the \\ Narrow-Line Seyfert 1 Galaxy Mrk~1044}

\correspondingauthor{Yair Krongold}
\email{yair@astro.unam.mx}

\author[0000-0001-6291-5239]{Y. Krongold}
\affiliation{Instituto de Astronom\'ia, Universidad Nacional Aut\'onoma de M\'exico,  Circuito Exterior, Ciudad Universitaria, Ciudad de M\'exico 04510, M\'exico}

\author[0000-0001-5948-8360]{A.L. Longinotti}
\affiliation{Instituto de Astronom\'ia, Universidad Nacional Aut\'onoma de M\'exico,  Circuito Exterior, Ciudad Universitaria, Ciudad de M\'exico 04510, M\'exico}

\author[0000-0001-5948-8360]{M. Santos-Lle\'o}
\affiliation{European Space Agency (ESA), European Space Astronomy Centre (ESAC), Camino Bajo del Castillo s/n, 28692 Villanueva de la Cañada, Madrid, Spain}

\author[0000-0002-4822-3559]{S. Mathur}
\affiliation{Department of Astronomy, The Ohio State University, 140 West 18th Avenue, Columbus, OH 43210, USA}
\affiliation{Center for Cosmology and Astroparticle Physics, 191 West Woodruff Avenue, Columbus, OH 43210, USA}

\author[0000-0001-6481-5397]{B. M. Peterson}
\affiliation{Department of Astronomy, The Ohio State University, 140 West 18th Avenue, Columbus, OH 43210, USA}
\affiliation{Center for Cosmology and Astroparticle Physics, 191 West Woodruff Avenue, Columbus, OH 43210, USA}

\author[0000-0002-6896-1364]{F. Nicastro}
\affiliation{Observatorio Astronomico di Roma-INAF, Via di Frascati 33, 1-00040 Monte Porzio Catone, RM, Italy}

\author[0000-0003-1880-1474]{A. Gupta}

\affiliation{Department of Astronomy, The Ohio State University, 140 West 18th Avenue, Columbus, OH 43210, USA}
\affiliation{Columbus State Community College, 550 E Spring St., Columbus, OH 43215, USA}

\author[0000-0002-6505-1349]{P. Rodr\'iguez-Pascual}
\affiliation{SERCO for the European Space Agency (ESA), European Space Astronomy Centre (ESAC), Camino Bajo del Castillo s/n, 28692 Villanueva de la Cañada, Madrid, Spain}

\author[0000-0002-0115-8374]{M. El\'ias-Ch\'avez}
\affiliation{Instituto Nacional de Astrof\'isica, \'Optica y Electr\'onica, Luis E. Erro 1, Tonantzintla, Puebla, C.P. 72840, Mexico}







\begin{abstract}
We present a detailed analysis of {\it XMM-Newton} X-ray spectra of the Narrow-Line Seyfert 1 galaxy Mrk~1044.  We find robust evidence for a multi-phase, ultra-fast outflow, traced by four separate components in the grating spectrum. One component has high column density and ionization state, and is outflowing at $\sim 0.15c$. The other three wind components have lower temperature,  lower column density, and have outflow velocities $\sim0.08c$. This wind structure is strikingly similar to that found in IRAS\,17020+4544, suggesting that stratified winds may be a common feature of ultra-fast outflows. Such structure is likely produced by fluid instabilities that form when the nuclear wind shocks the ambient medium. We show that in an energy-driven wind scenario, the wind in Mrk 1044 might carry enough energy to produce significant feedback on its host galaxy. We further discuss the implications of the presence of a  fast wind in yet  another NLS1 galaxy with high Eddington ratio.

\end{abstract}



\section{Introduction} \label{sec:intro}

Outflows of copious amounts of gas are a common feature observed in many bands of the electromagnetic spectra of active galactic nuclei (AGNs).
The ``warm absorbers'' that are observed in 
the X-ray spectra of about 50\% of Seyfert galaxies provide a detailed view of the ionized gas  that is outflowing at velocities of $10^{2-3}\,\kms$ in the nuclear regions
\cite[][and references therein]{Laha2014,Laha2021}. 

Much faster winds, the ``ultra-fast outflows,'' or UFOs, with $v_{\rm out} > 10,000\,\kms$, were initially discovered two decades ago  \cite[e.g.][]{Chartas2002,Pounds2003,Reeves2003, Dasgupta2005,Obrien2005}. Systematic studies of these systems show a detection fraction in 30--40\% of nearby AGNs and indicate they are likely launched at accretion disk scales, \citep{Tombesi2010,Gofford2013, Tombesi2013}.
These winds show mass outflow rates of around
$0.01$--$1M_{\odot}\,{\rm yr}^{-1}$, kinetic energies around $10^{42-45}\,\ergsec$, and are consistent with theoretical predictions of black hole feedback models
\citep{King2010}.  

The effect of  AGN winds on the larger-scale environment, commonly referred to as ``AGN feedback,''  may have a profound impact on the evolution of the host-galaxy properties 
\citep{DiMatteo2005, Hopkins2010}. Indeed, quasar feedback is  postulated as the mechanism capable of regulating the relationships between supermassive black holes and their host galaxies that is required by simulations in order to reproduce the galaxy luminosity function at high luminosities \citep[see][]{Bower2006,Kormendy2013}. 
The discovery of sub-relativistic X-ray winds in very bright quasars \citep[such as PDS 456,][]{Reeves2003,Nardini2015} and the relation of the inner X-ray wind to galaxy-scale outflows 
(\citealt[][]{Tombesi2015}, but see revised estimates of energetics by \citealt[][]{Veilleux2017} ; \citealt{ 
Feruglio2015, Longinotti2018, Mizumoto2019,Bischetti2019, Marasco2020}) supports the role of accretion disk winds as potential feedback agents. This feedback is expected to play a major role in the early Universe, during the peaks of star formation and black hole accretion in the Universe. However, due to the intrinsic difficulty in detecting these systems, only a few examples of a connection between a sub-pc outflow and a large scale wind  are known at z$>$1 \citep[e.g.][]{Feruglio2017, Chartas2020}.  

Narrow-line Seyfert 1 (NLS1) galaxies constitute a particular sub-class of AGNs that is defined by an 
\hb\ line width ${\rm FWHM} < 2000\,\kms$ and
unusually prominent \feii\ lines \citep{Osterbrock1985}.
They are also characterized by extreme X-ray properties 
\citep{Gallo2018} that include  
strong soft X-ray emission characterized by steep photon indices, a high degree of continuum variability, and high Eddington ratios. 
Recent serendipitous analyses of X-ray grating spectra of NLS1s have shown that these sources seem to preferentially host 
UFOs with a very rich stratified structure 
\citep{Gupta2013,Gupta2015,Longinotti2015, Parker2017,Reeves2018,Reeves2019}.

Considering that the archetypical fast wind-quasar PDS~456 \citep{Reeves2003,Nardini2015} also shows a dependence of the wind parameters on the source luminosity and accretion rate 
\citep{Matzeu2017}, the occurrence of ultra-fast outflows seems to be related to phenomena taking place in highly accreting sources. A straightforward explanation would be that  the wind is launched from the accretion disk via radiatively driven processes \citep{Pinto2018}. However,  recent results suggest that radiative pressure cannot accelerate UFOs to $\sim$0.1c because when special relativistic effects are taken into account the absorption cross section of the atoms is greatly reduced. This implies that the pressure exerted by photons on the fast-escaping gas particles is also greatly reduced  \citep[][]{Luminari2021}. As such, the exact mechanism behind the launching and acceleration of UFOs remains a mystery.

Mrk~1044 is a nearby NLS1 at $z = 0.016$ with $L_{\rm BOL} =1.4\times10^{44}\,\ergsec$ \citep{Grupe2010}. It is
 characterized  by a very high Eddington ratio, 
 with reported values ranging from 
 $\dot{m} \approx 0.59$ \citep{Grupe2010} to $\dot{m} \approx 16$ \citep{Du2015}. Mrk~1044 is a barred spiral galaxy, relatively undisturbed \citep{Deo2006,Powell2018} with no signs of interaction
 \citep{Koulouridis2006}, and a modest star-formation rate ($L_{\rm FIR}=3\times10^9\,L_\odot$; \citealt{Bertram2007}). It has a small black hole mass 
$M_{\rm BH}=2.8\times10^6\,M_{\odot}$, based on 
reverberation-mapping measurements (\citealt{Hu2015,Du2019}, though see
\S\ref{section:implications}). Despite being one of the X-ray brightest AGN in the local Universe, it has not been extensively observed by X-ray facilities until relatively recently. 
\cite{Dewangen2007}  reported on a short {\xmm}  exposure, which highlighted a strong featureless soft excess and complex temporal behavior. 
An extensive analysis of a deeper {\xmm} observation coupled with {\it NuSTAR} and 
{\it Neil Gehrels Swift Observatory} data \citep{Mallick2018} concluded that  the excess emission observed in the soft and hard X-ray bands can be explained by relativistic Compton reflection from a high-density accretion disk.
\cite{Mallick2018} also tentatively reported the presence of a soft X-ray fast outflow. Comparison with this work is presented in \S \ref{sec:disc}.

An intrinsic outflow has been observed in the UV band of Mrk~1044 with a velocity $v \approx -1150\,\kms$ \citep{Fields2005a,Fields2005b} based on spectra obtained with 
the {\it Far Ultraviolet Spectroscopic Explorer (FUSE)}
and the {\it Hubble Space Telescope (HST)}. By comparing the \ovi, \civ, and \nv\ column densities with that of \hi, they
conclude that the metallicity of the outflow is supersolar.

This paper presents the first detailed spectroscopic analysis of the high resolution RGS data of Mrk~1044. 

\section{Observations and Data Reduction}

 XMM–Newton observed Mrk 1044 on 2013 January 27 for a total duration of 133ks (OBSID 0695290101, PI M. Santos-Lleó). The XMM–Newton EPIC-pn camera was set to operate in Small Window mode. The data were processed with SAS 18.0.0. The raw event file was processed with the SAS task EPPROC. Events were subsequently filtered with PATTERN $\leq$4 (therefore including single and double events) and FLAG=0. The Good Time Interval (GTI) was created via standard procedure by filtering the 10-12 keV light curve of the field of view with a rate of 2 cts s$^{-1}$ to exclude background flares. Photon pile up was checked with the EPATPLOT task and found not to be an issue. The net exposure in the resulting event list is 91 ks. 

Source and background spectra were extracted from a circular region of 40 and 53 arc sec radius, respectively, and response matrices were produced with the SAS tasks RMFGEN and ARFGEN. The source spectrum was rebinned with the task SPECGROUP to have a minimum number of 25 counts in each spectral channel  and in order to avoid oversampling of the instrumental resolution by a factor larger than 3. We note that the background in the pn spectrum dominates over the source spectrum for energies above $\sim$8keV. 
 
\begin{figure}
	\includegraphics[width=1.\columnwidth]{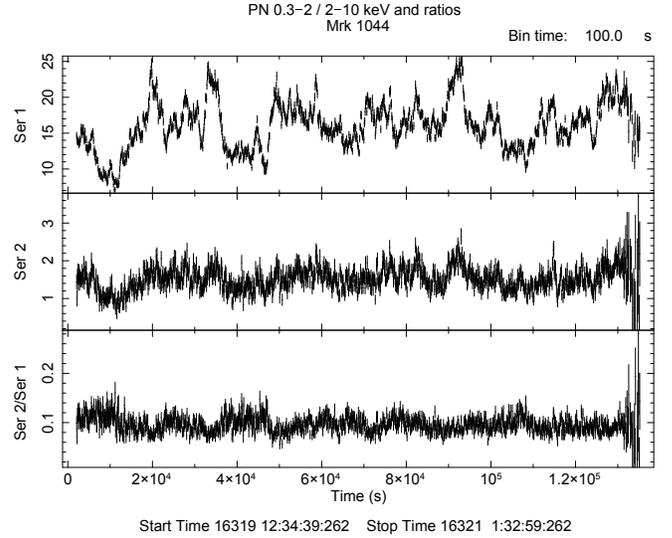}
    \caption{Light curve of Mrk~1044 extracted from the  {\xmm} observation. From top to bottom: background subtracted source counts in 0.3--2 keV, 2--10 keV and hardness ratio. }
    \label{fig:lcurve}
\end{figure}

\begin{figure}
	\includegraphics[width=1.\columnwidth, angle=0]{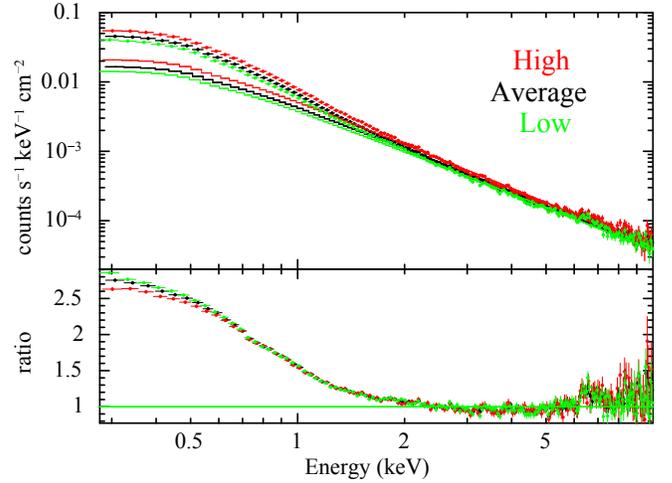}
    \caption{Upper panel: EPIC-pn spectra extracted form the average, low, and high states of the light curve. The spectra have been modelled in the 2-10 keV region with a simple powerlaw (solid lines in the plot). Lower panel: ratio between the data and the hard band powerlaw model. We do not find strong changes in the spectral shape between the high and low flux states.}
    \label{fig:3spec}
\end{figure}

The background-subtracted and deadtime-corrected light curve extracted from the pn counts with the task EPICLCCORR reveals strong variability during the XMM–Newton observation on timescales of few ks (see Fig.~\ref{fig:lcurve}). The hardness ratio, plotted in the bottom panel of Figure~\ref{fig:lcurve}, indicates that the behavior of the source flux is similar in the hard and soft bands. As a first-order check for spectral variability between the flaring and quiescent portions of the light curve, we assumed a threshold count rate of 17 cts s$^{-1}$. Consequently, events were selected above and below this threshold. Spectra extracted from the “high” and “low” states following the same procedure described above for the integrated spectrum, do not show any obvious change in the spectral shape. This can be clearly observed in Figure~\ref{fig:3spec}, where the average, high, and low state spectra are presented against a simple hard-band powerlaw model. 

Data obtained by the RGS instrument were processed through the standard RGSPROC  tool and the combined spectrum of RGS1 and RGS2 cameras was obtained via the standard RGSCOMBINE tool.  In addition to the spectrum from the full exposure of 130 ks, spectra from the “high” and “low” periods were extracted following the procedure applied to the pn data. RGS spectra were not grouped during the spectral analysis.  The total exposure accumulated by the RGS in the high and low states is 50 ks and 80 ks, respectively. 

Chi-squared statistics was applied when fitting the CCD data and C-statistics \citep[][]{Cash1979} was applied for the grating spectral analysis. 
A detailed temporal analysis of the EPIC data was published by \cite{Mallick2018}. For the purpose of this study, we have compared the high and low EPIC and RGS spectra, applying the same spectral analysis to both of them. We checked for variations in the properties of the X-ray wind detected in the entire spectrum (see \S \ref{sec:ufo}). Despite the large flux variations (a factor $\sim$3) we find no changes in the wind properties between the high and low states. Therefore, we will discuss only results based on the integrated spectrum from the full exposure throughout the rest of this paper. In the following, errors are quoted at the 1$\sigma$ level.

\begin{figure*}
	\includegraphics[width=.5\linewidth]{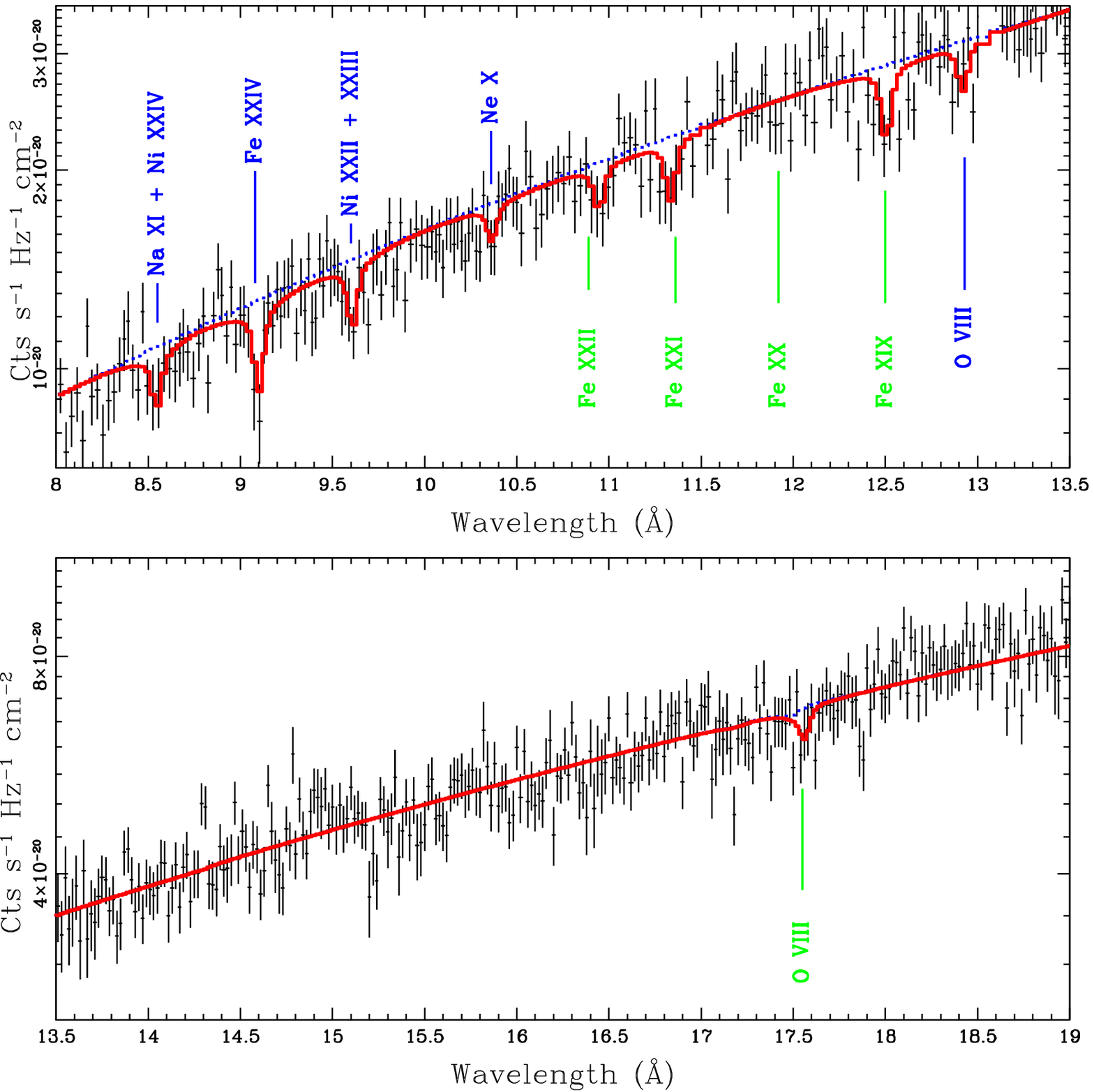}
     \includegraphics[width=.5\linewidth]{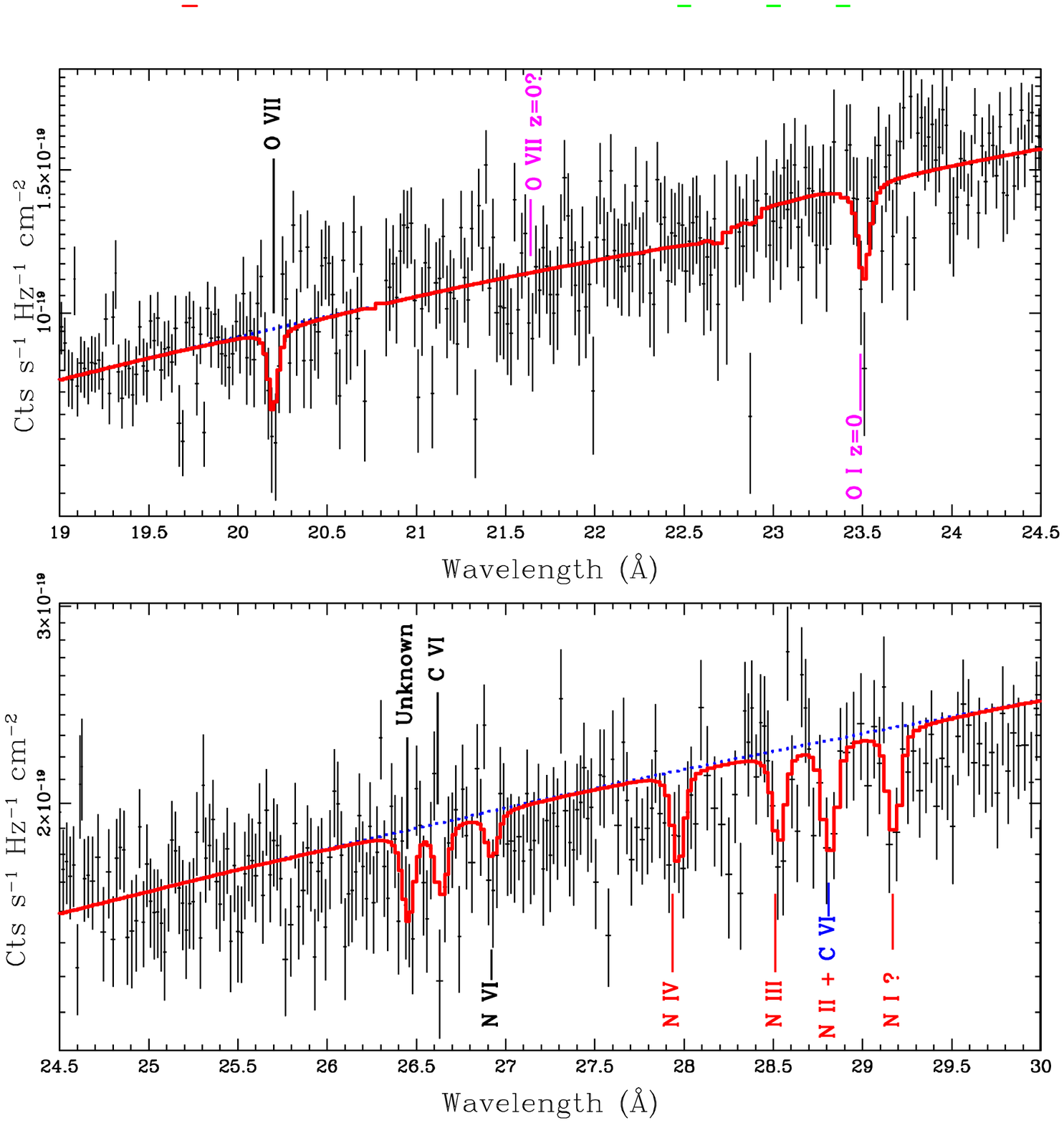}
    \caption{ RGS spectrum in the 8-30 \AA\ range,  fitted with a model including a powerlaw continuum and 16 Gaussians. Main observed/expected absorption lines by the CGM of our Galaxy at z=0 are marked in magenta. Line identifications were made by matching this phenomenological fit to the
physical model in Table \ref{tab:wa_par}.  The absorption features coincident with the outflow velocity of UFO1 are labeled in blue, and those with the velocity of UFO2 are marked in green. Features consistent with the outflow velocity of UFO3 and UFO4  are shown in red and black, respectively. The blue dotted line shows the model without the Gaussians. The spectrum was binned by up to four instrumental channels for display purposes only.}
    \label{fig:rgs1}
\end{figure*}

\begin{figure}
\centering
	\includegraphics[width=\columnwidth]{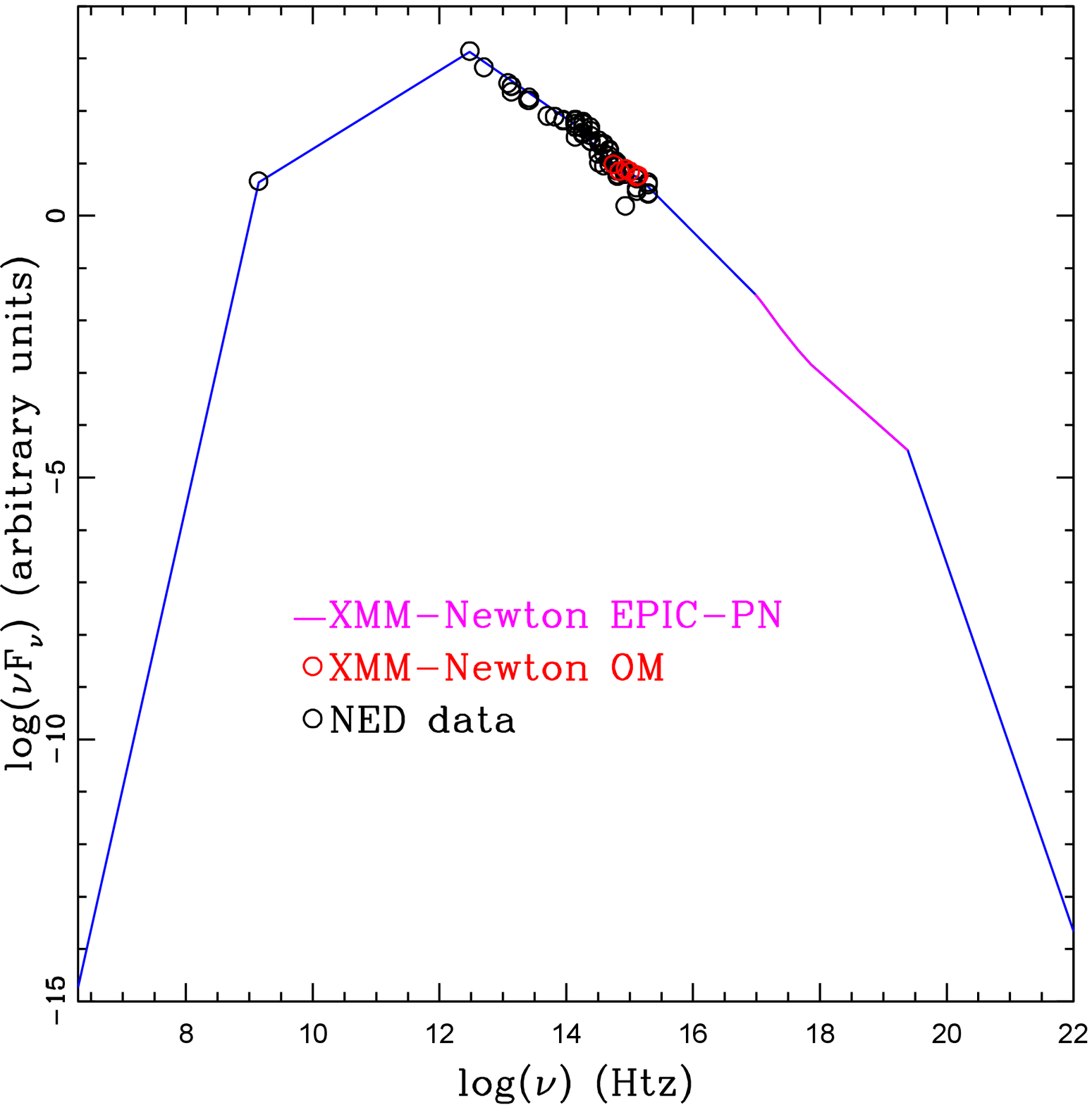}
\caption{Spectral energy distribution of Mrk~1044 used to model the ultra-fast outflow wind in this source. Data were obtained from the NASA Extragalactic Database (black points) and from the {\it XMM--Newton} observation discussed here (red points: Optical Monitor, green line: EPIC-PN best fit model).}
    \label{fig:sed}
\end{figure}

\section{Spectral Analysis of the High Resolution Data of Mrk~1044\label{sec:ana}}

We carried out the analysis of Mrk~1044 RGS data using the unbinned spectrum, making use of  \cite{Cash1979} statistics to perform the spectral fitting in the software package {\tt Xspec} (V.\ 12.11), and subtracting the background spectrum from the source spectrum.\footnote{ According to the Xspec manual, the use of Cash statistics with a background spectrum with Poisson statistics can produce a bias in the fit. This is expected only for weak sources and small number of counts in the background spectrum. Although this is not the case of our source, we double-checked our fits running them on the data binned by a factor of 10. We find similar results to those on the unbinned data, confirming that our results are not affected by this bias.    
 }

\subsection{Modeling the Spectrum with a simple phenomenological model \label{sec:powerlaw}}

We started by fitting the continuum in the 8--30\,\AA\ (0.4--1.5\,keV) range with a phenomenological model consisting of a powerlaw attenuated by Galactic absorption, assuming  a column density  $N_{\rm H} = 3.6\times10^{20}\,\pcm$ \citep{Willingale2013}, and
using the {\tt Xspec} model {\tt TBabs}. Although this is not a physically motivated  model, it represents the most simple description of the continuum to look for the presence of absorption or emission features. The model gives a reasonable fit to the data, with a $C$ statistic of 2541.5 for 2162 degrees of freedom (d.o.f.). The photon index of the powerlaw is $\Gamma=2.66\pm 0.01$, typical of the high values obtained in the soft X-ray band for NLS1s.

\begin{deluxetable*}{rrcccc}
\tablecaption{List of absorption lines individually detected above a 2.5$\sigma$ threshold.\label{tab:lines}}
\tablewidth{0pt}
\tablehead{
\colhead{Obs. $\lambda$} & \colhead{Intensity} & \colhead{$\Delta$C$_{stat}$} & \colhead{Sign.} & \colhead{Line ID} & \colhead{Component}\\
\colhead{\AA} & \colhead{$10^{-5}$ ph cm$^{-2}$ s$^{-1}$} & \colhead{}  & \colhead{$\sigma$} & \colhead{} & \colhead{}
}
\startdata
8.55$^{+0.03}_{-0.01}$ & 1.15$^{+0.12}_{-0.38}$ & 8.0 & 3.0 & \naxi + \nixxiv & UFO1 \\
9.09$\pm{0.01}$ & 1.72$\pm{0.38}$ & 17.6 & 4.5 & \fexxiv & UFO1 \\
9.61$^{+0.01}_{-0.02}$ & 1.30$^{+0.34}_{-0.30}$ & 15.0 & 4.3 & \nixxii + \nixxiii & UFO1 \\
10.36$^{+0.01}_{-0.03}$ & 0.79$\pm{0.30}$ & 6.1 & 2.6 & \nex & UFO1 \\
10.94$\pm{0.04}$ & 1.04$\pm{0.41}$ & 5.0 & 2.5 & \fexxii & UFO2 \\
11.35$^{+0.03}_{-0.01}$ & 1.33$^{+0.45}_{-0.38}$ & 9.7 & 3.5 & \fexx & UFO2 \\
12.50$^{+0.01}_{-0.05}$ & 1.61$\pm{0.41}$ & 13.6 & 3.9 & \fexix & UFO2 \\
12.93$^{+0.04}_{-0.01}$ & 1.12$\pm{0.43}$ & 6.2 & 2.6 & \oviii ? & UFO1 \\
17.57$^{+0.06}_{-0.05}$ & 0.89$\pm{0.35}$ & 5.1 & 2.5 & \oviii  & UFO2 \\
20.20$\pm{0.01}$ & 1.73$\pm{0.56}$ & 8.1 & 3.1 & \oviii  & UFO4 \\
26.45$\pm{0.02}$ & 1.79$\pm{0.63}$ & 7.6 & 2.8 & ? & ? \\
26.64$\pm{0.01}$ & 1.61$\pm{0.61}$ & 6.8 & 2.6 & \cvi & UFO4 \\
27.96$\pm{0.03}$ & 2.00$\pm{0.58}$ & 10.3 & 3.4 & \niv & UFO3 \\
28.53$\pm{0.01}$ & 2.05$\pm{0.63}$ & 9.2 & 3.3 & \niii & UFO3 \\
28.81$\pm{0.01}$ & 3.25$\pm{0.65}$ & 20.6 & 5.0 & \cvi & UFO1 \\
29.17$\pm{0.01}$ & 2.55$\pm{0.69}$ & 12.4 & 3.7 & \niti ? & UFO3 \\
\enddata
\tablecomments{Identifications were made by matching this phenomenological fit to the
physical model in Table \ref{tab:wa_par}}
\end{deluxetable*}

We did not find any evidence of residuals pointing to the presence of emission features in the spectrum. However, the data show negative residuals throughout the spectral range, suggestive of absorption lines by ionized gas. We performed a blind search for individual absorption lines by adding narrow Gaussian with negative intensity to our simple powerlaw model, over the whole spectral range. The line widths were fixed to  0.1 eV and their positions and intensities were left free to vary. All the lines detected above a 2.5$\sigma$ threshold are reported in Table \ref{tab:lines}. The improvement in C-statistics is measured for the addition of 2 d.o.f. per line. The significance of the lines was computed by dividing the intensity of the line over the positive 1$\sigma$ error. Figure \ref{fig:rgs1} presents our phenomenological best fit model consisting of a simple powerlaw and 16 Gaussians. Several of these features are highly significant (10 lines have $\sigma\geq$3 and three have $\sigma\geq$4), showing that the presence of absorbing gas is required by the data. However, the line positions do not match the common transitions found in warm absorbers, which typically have outflow velocities up to few thousand kilometers per second in the rest frame of the source. A clear identification of the lines was only possible with the aid of physical models to account for the absorption. Identifications in Figure \ref{fig:rgs1} and Table \ref{tab:lines} are based on these models. The outflow velocity of these absorbers puts them in the range for UFOs, as we show in the next section.

\subsection{A Physical model for the UFOs in Mrk 1044 \label{sec:ufo}}

To produce a physical description of the absorption in the RGS spectrum of Mrk~1044,  we opted to apply a self-consistent model. For this purpose, we used the photoionization code {\tt PHASE} \citep{Krongold2003}, which assumes a slab of ionized gas in a plane-parallel geometry described by four parameters\footnote{A traditional fifth parameter, the covering factor in the line of sight, was not considered since the models were consistent with a value of unity.}: 
\begin{enumerate}
\item The ionization parameter\footnote{The dimensionless ionization parameter is defined as 
\begin{equation}
U =  \frac{Q}{4\pi R^2 c n_e}, \nonumber
\end{equation}
where $Q$ is the rate at which the source emits ionizing photons (i.e.. photons per second), $c$ is the speed of light, 
and $n_e$ and $R$ are the electron density and the distance of the gas from the X-ray source, respectively.};
\item The equivalent hydrogen column density of the slab $N_{\rm H}$;
\item The outflow velocity of the wind $v_{\rm out}$; and
\item The microturbulence velocity of the medium.  Given that the absorption features in the spectrum are narrow and unresolved, we fix the microturbulence velocity to $10\,\kms$, however, we note that the exact value adopted for this parameter has negligible effect on the fit. 
\end{enumerate}

\begin{figure*}
	\includegraphics[width=.49\linewidth]{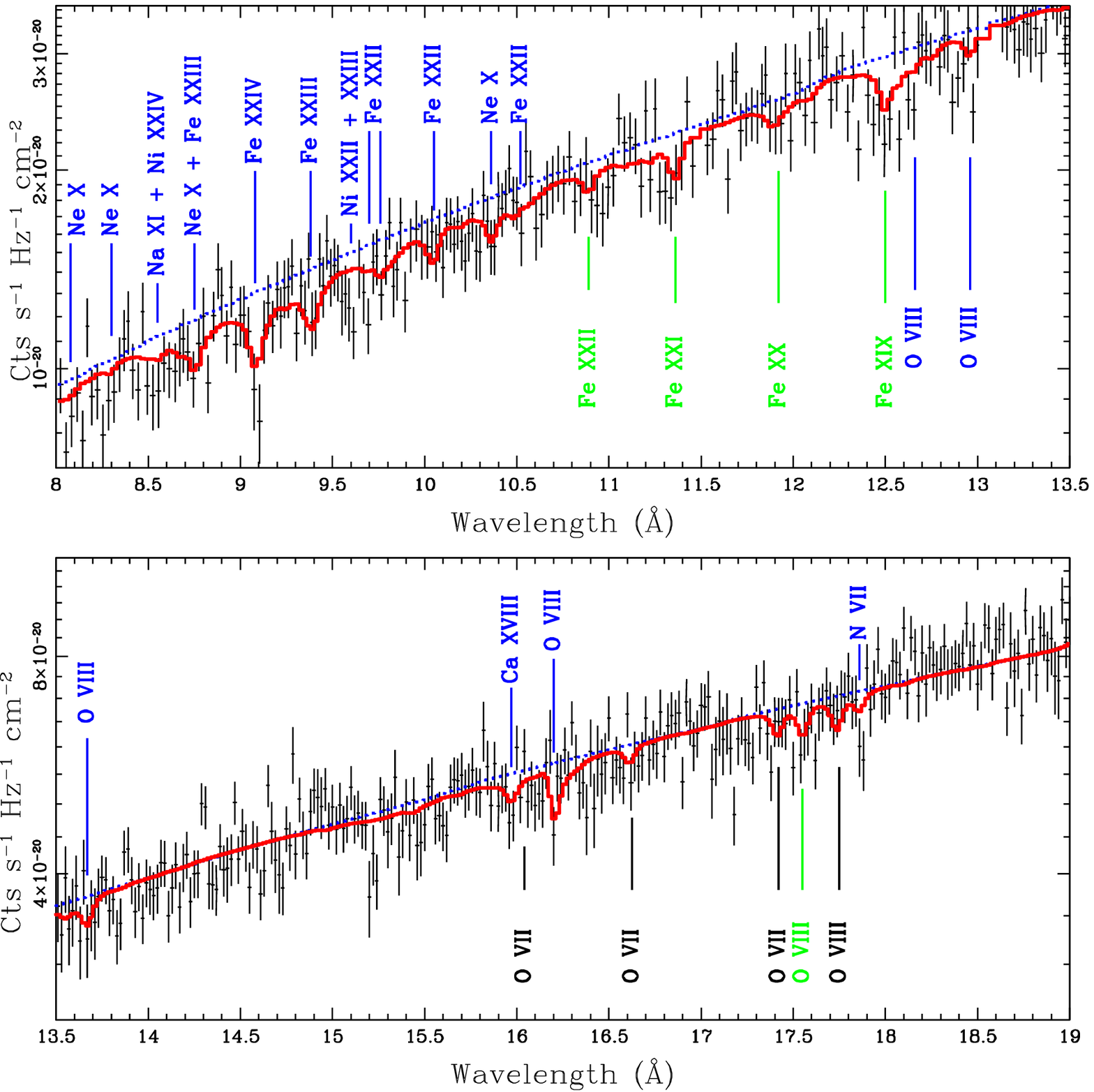}
    \includegraphics[width=.49\linewidth]{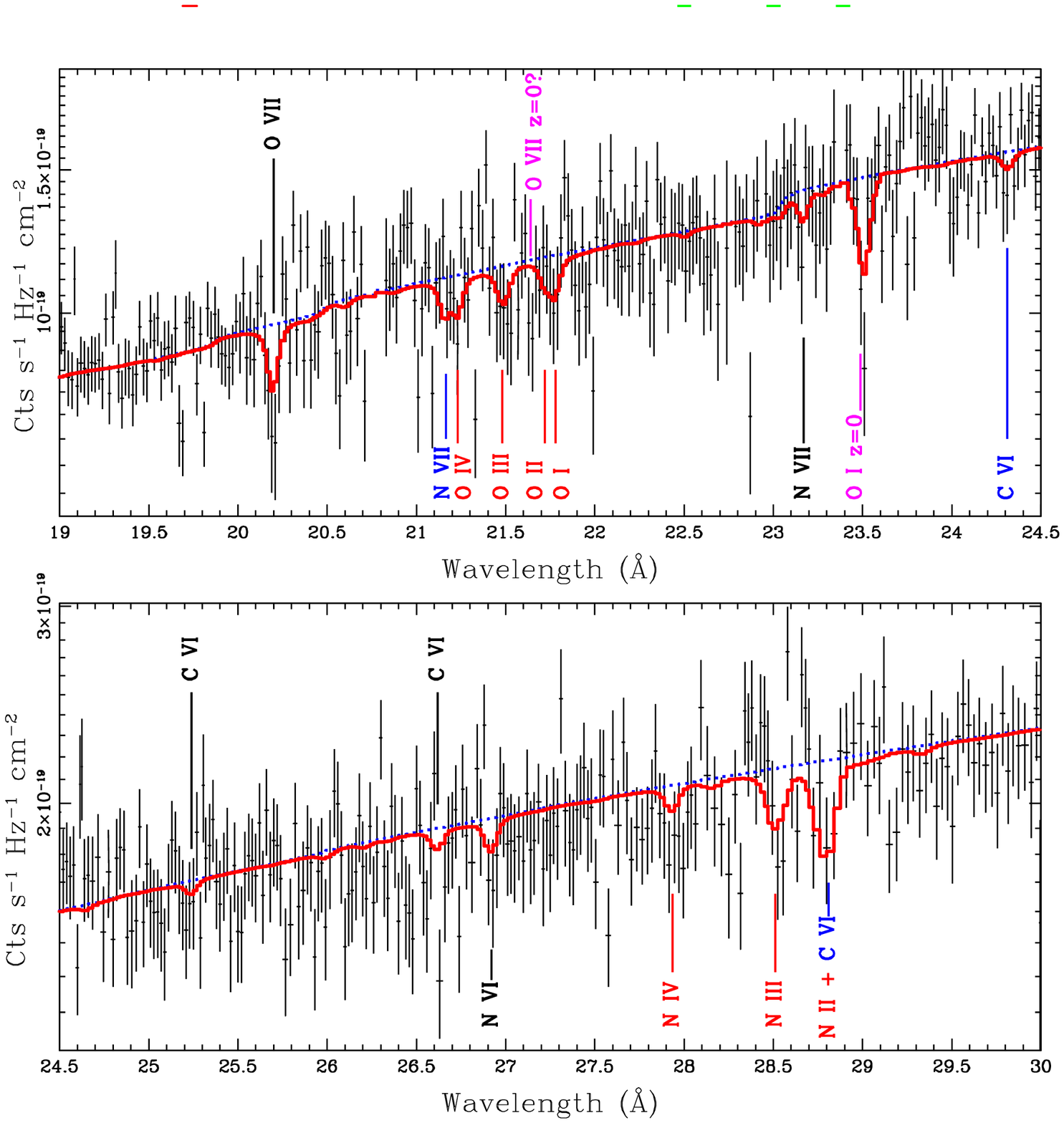}

    \caption{RGS spectrum in the 8--30\,\AA\ range,  fitted with a model including a powerlaw continuum attenuated by four UFO components. The main absorption lines produced by UFO1 are labeled in blue, and those by UFO2 are marked in green. Features by UFO3 and UFO4  are shown in red and black, respectively. Expected absorption lines by the CGM of our Galaxy at $z=0$ are marked in magenta. The blue dotted line shows the model without the contribution of bound-bound absorption. The spectrum was binned by up to four instrumental channels for display purposes only.}
    \label{fig:rgs4}
\end{figure*}

To calculate the ionization balance produced by the impinging ionizing radiation of the source, we constructed the spectral energy distribution (SED) for Mrk~1044 using data from the NASA Extragalactic Database 
and from the {\it XMM--Newton} spectrum discussed here  
(see Fig. \ref{fig:sed}).

\subsection{Evaluating the significance of the UFO components \label{sec:significance}}

While fitting the data with the PHASE components, we evaluated the relative quality of a model including an UFO using two different methods. 

First, we used the Akaike Information Criterion values \citep[][]{Akaike1974}, as done by \citet[][]{Hebbar2019}. These values are defined as\footnote{We note that since we are using the unbinned RGS data, we have a large number of d.o.f. in our models, and we do not require the correction of the Akaike values for small sample sizes.}: 
 \begin{equation}
AIC = 2k + C_{stat}
\end{equation}
where $k$ is the number of parameters and $C_{stat}$ is the value of the C statistic for the best fit with a given model. Then, the relative likelihood that the model including the UFO component is better than the model without it is given by the quantity exp((AIC(UFO)- AIC(No-UFO))/2). As noted by \citet[][]{Hebbar2019}, the inverse of this quantity represents the factor by which the more complex model (containing the UFO) is preferred over the simpler model.

The second method to estimate the statistical significance of the UFO components consisted in running Monte Carlo simulations. To do this, we followed the procedure in
\citet[][hereafter L15]{Longinotti2015}. As a baseline model for these simulations, we employed the phenomenological continuum  model described in \S \ref{sec:powerlaw} (consisting in a powerlaw with Galactic absorption). We include any previous UFO component already in the model, as done in the actual spectral fitting (see \S \ref{sec:results_rgs}). For example, we included in the simulations only the continuum model to test the significance of one UFO, we included continuum+UFO1 for testing the significance of the second UFO, etc. 
For each UFO component, we produced 1000 synthetic spectra with the {\tt XSPEC} package, each with the same photon statistics as the real data set. Following L15, we recorded the improvement in $C$-statistics by adding a new {\tt PHASE} component to the baseline model. The final distribution of $\Delta C_{sim}$ in the simulations was then fit assuming a Gaussian profile.
\footnote{The actual distribution of $\Delta C_{sim}$ does not follow an exact Gaussian profile, but assuming such a profile represents a very conservative way to estimate the significance of the fitted components.}. 
The significance of each UFO component was calculated by dividing the $\Delta C_{Stat}$ of the best fit model in the real data, over the 1$\sigma$ range of the Gaussian model to the distribution of $\Delta C_{sim}$ in the simulations.

\subsection{Results to the fit of the RGS data with the PHASE components \label{sec:results_rgs}}
We started by including one {\tt PHASE} component to our model, thus we fit the data with a powerlaw attenuated by a slab of photoionized gas (with the ionization parameter, the column density, and the outflow velocity of the absorber free to vary), and further attenuated by the gas in our Galaxy. This model provides a large improvement to the fit, with  $\Delta C = 81/3$ d.o.f.  (the C-statistic of the fit with a simple powerlaw was 2541.5 for 2162 d.o.f.), pointing to the presence of a highly significant absorber in the line of sight (see below).  We note that this absorber (hereafter UFO1) has a very large outflow velocity, $v_{\rm out}\approx0.15c$ (see Table \ref{tab:wa_par}). Thus, in addition to its high column density ($N_{\rm H} \approx 10^{23}\,\pcm$), the absorption lines are strongly shifted to the blue with respect to the rest frame of Mrk~1044, explaining why identification of these features was not possible by simple visual inspection of the spectrum. The ionization state of UFO1 is also quite high 
($\log U \approx 2.2$), producing absorption by ions such as 
\cvi, \nvii, \oviii, \nex, and \fexxii--\fexxiv, among others,
but not high enough to produce \fexxv--\fexxvi,
as observed in other UFOs \citep[e.g.,][]{Parker2018}.
The significance of UFO1 was evaluated as described in \S \ref{sec:significance}. It turned out to be much greater than $10\sigma$, using Monte Carlo simulations. The comparison of AIC statistics (Table \ref{tab:wa_par}) also indicates a very large significance: The model including UFO1 is preferred by a factor of 2$\times10^{16}$ over the model considering only the continuum.

\begin{table*}

\caption{Best fit parameters of the phenomenological continuum and the multiphase ultra-fast wind detected in the RGS spectrum of Mrk~1044.}
\centering
\begin{tabular}{c c c c c c c c}
\hline
\multicolumn{8}{|c|}{Single Powerlaw Continuum Parameters} \\
\hline
$\Gamma$  &     &  \multicolumn{2}{c}{Norm.($10^{-3}$ ph keV cm$^{-2}$ s$^{-1}$)}  & & AIC &  &\\
\hline
$2.82\pm 0.01$ & & \multicolumn{2}{c}{$8.0\pm 0.1$}  & & 2545.5 & &\\ 
\\
\hline
\multicolumn{8}{|c|}{Wind Parameters} \\
\hline
 Wind Phase  & $\log U$       &  $\log N_{\rm H}$     & $v_{\rm out}$   &   Statistics & AIC & Factor &  Significance  \\
    -  & 
    &  (cm$^{-2}$) &        (km s$^{-1}$) &  $\Delta C$/$\Delta\nu$   &  \\
    \hline
 UFO1  &  $2.2\pm0.1$   &  $22.9_{-0.1}^{+0.2}$ & $44800_{-70}^{+130}$ &  81/3  & 2470.5 & 2$\times10^{16}$&  $>>10\sigma$  \\
           
UFO2   &  $1.6\pm0.1$  &  $21.2_{-0.024}^{+0.25}$ & $26084_{-240}^{+90}$ &  21/3 & 2455.5 & 1808 & $4\sigma$ \\ 

UFO3   &   $-2.7\pm0.1$ &  $20.6\pm0.1$  &  $23670_{-270}^{+180}$     &  36/3 & 2425.5 & 3$\times10^{6}$ & $6\sigma$  \\ 

UFO4    &  $-0.2 \pm0.1$ & $20.3\pm0.1$ &  $23430_{-110}^{+190}$   &  17/3 & 2414.5 & 245 & $3\sigma$   \\
\\
    \hline
\multicolumn{8}{|c|}{Final Statistic} \\
    \hline
\multicolumn{8}{c}{C$_{stat}$/$\nu$ \ \ \ \ \ 2386.5/2150 } \\
\hline\hline
\end{tabular}
\label{tab:wa_par}
\end{table*}

Although the presence of UFO1 accounts for a large part of the residuals to a powerlaw in the spectrum, strong residuals remain (including the Gaussian fits not identified with this UFO, see Fig.~\ref{fig:rgs1}). Therefore, we continued fitting the data by iteratively adding additional {\tt PHASE} components to the model (as we did with UFO1, for each component, the ionization parameter, column density, and outflow velocity were free parameters). Three more statistically significant {\tt PHASE} components (see below),
each with $\Delta  C > 15$ for the three new free parameters, were required. This components are hereafter designated
UFO2, UFO3, and UFO4. Table \ref{tab:wa_par} presents the results of our best fit, including the continuum and four UFO components. Figure~\ref{fig:rgs4} compares the best fit with the spectrum.  

Statistical significances for UFO2, UFO3, and UFO4 were evaluated as described in \S \ref{sec:significance}. Using Monte Carlo simulations, they are $\sim 4\sigma$ for UFO2,  $\sim 6\sigma$ for UFO3, and  $\sim 3\sigma$ for UFO4. AIC statistics comparison indicates that UFO2, UFO3 and UFO4 improve the fit by factors~$\approx$~1800,~3$\times10^{6}$,~and 250, respectively (see Table \ref{tab:wa_par}).

The new UFO components have roughly half the speed of UFO1 ($v \approx 0.08c$), and at least 50 times lower column density. UFO2 is highly ionized 
and absorption by \oviii\ and \fexix--\fexii\ is detected
(green labels in Fig.~\ref{fig:rgs4}). On the other hand, UFO3
has very low ionization, with dominant ions \oi--\oiv\ 
and \nii--\niv\  (red labels in Fig.~\ref{fig:rgs4}). Finally
UFO4 has an intermediate level of ionization, with features due to \ovii--\oviii\ and \nvi--\nvii. It is noteworthy  that UFO3 and UFO4 have strikingly similar velocities, fully consistent within the errors, pointing to a common origin for these two components.

 Our final fit still leaves some residuals, particularly near 26.45 and 29.17\AA\ (see Table \ref{tab:lines}), where Gaussians with significance 2.8 and 3.7$\sigma$ were required.  The line at 29.17\AA\ is consistent with absorption by \niti\ at the same velocity of UFO3, however this component does not have enough neutral material to produce this feature. This indicate that the structure of this UFO component might be more complex than the simple absorbing slab model presented here. The feature at 26.45\AA\ cannot be accounted for with the addition of a further significant UFO component or with the presence of a Galactic line at z=0, and thus, it remains unidentified in our analysis.

\section{Analysis of the Low-Resolution EPIC-pn data \label{sec:pn}}

Motivated by our results on the wind found with the RGS data, we modelled the EPIC-pn data with two goals in mind. First, 
 we wanted to test the presence of the UFO components found in the RGS data. Second, we  wanted to search for  higher-ionization absorption by \fexxv\ and \fexxvi\ that might coexist with the presence of UFO1, whose ionization range extends up to \fexxiv.

\citet[][]{Mallick2018} found that the broad-band spectrum of Mrk~1044 can be well described by a model consisting of  relativistic reflection from a high-density accretion disk
with a broken powerlaw emissivity profile. Given that this model represents a physically motivated description of the continuum, we follow their analysis. The EPIC-pn data were modelled with a primary continuum component represented by a simple powerlaw, a relativistically blurred reflection component to account for 
the accretion disk reflection (modelled with the high-density model RELXILLD; \citealt[][]{Garcia2016}), and a distant reflector to account for the narrow FeK$\alpha$ emission line (XILLVERD model; \citealt[][]{Garcia2016}). All these components were further attenuated by Galactic absorption.

A description of the parameters for the XILLVERD and RELXILLD models is presented in Table \ref{tab:epic}. We set the spectral index of these components to be the same as that of the direct continuum. The RELXILLD model was included in the fit with the rest of its parameters free, with the exception of the outer radius of the disk, which was fixed to 1000r$_g$ (where r$_g$ = GM${\rm_{BH}/c^2}$).
For the XILLVERD component, we fix the ionization parameter to 0, assuming neutral gas for the distant reflector. We also fixed the density to $10^{15}$ cm$^{-3}$ (the lowest value allowed by the model), and the inclination angle to 60$^{\rm o}$. 
In this way, the broad-band continuum was fit with 14 free parameters, two of the primary continuum source, 10 of the relativistic reflector and two of the distant reflector (Table \ref{tab:epic}). 

\begin{figure}
 \centering
	\includegraphics[width=1.0\columnwidth]{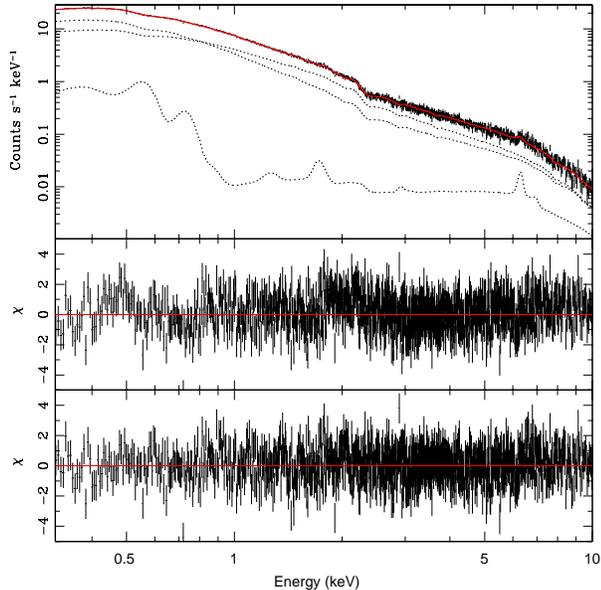}
\caption{Model and residuals between the EPIC-pn spectrum and the model discussed in \S \ref{sec:pn} in  the 0.3--10\,keV range. Top panel: Data and best fit model. Middle panel: Residuals to a continuum model consisting of a primary continuum source plus relativistic and distant reflection. Lower panel: Residuals to the best fit model including continuum reflection components plus two UFO components. The residuals are plotted in units of the standard deviation.}
    \label{fig:delchi}
\end{figure}

As in \citet[][]{Mallick2018}, we ignored the spectral range between 1.8-2.5 keV because of the instrumental edges produced by Si at 1.8 keV and Au at 2.2 keV \citep[e.g.][]{Marinucci2014,Matt2014}. This model provides a relatively good representation of the data, with $\chi^2=1059.5$ for 947 d.o.f. The residuals to this fit are preseneted in the middle panel of Figure \ref{fig:delchi}.

\begin{table*}
\centering
\caption{Best-fit parameters for the physically motivated continuum and the multiphase ultra-fast wind detected in the EPIC-pn spectrum of Mrk~1044.}
\begin{tabular}{l  c  l  }
\hline
\multicolumn{3}{|c|}{Primary Continuum Source} \\
\hline
$\Gamma$  &   $2.39\pm 0.01$   &  Photon index \\
N$_{pc}$  ($10^{-3}$ ph keV cm$^{-2}$ s$^{-1}$) & $4.6\pm0.3$ &  Normalisation \\ 
\hline
\multicolumn{3}{|c|}{Relativistic Reflection (RELXILLD)} \\
\hline
q$_{in}$  &   $8.17\pm 0.24$ & Inner emissivity index  \\
q$_{out}$  &   $1.72^{+0.34}_{-0.25}$ & Inner emissivity index  \\
R$_{br}$ (r$_g$)  &   $6.5^{+3.5}_{-0.3}$ & Break disk radius  \\
a  &   $0.994^{+0.002}_{-0.008}$ & SMBH spin  \\
$\theta^o$  &   $46.5^{+1.8}_{-7.3}$ & disk inclination angle  \\
R$_{in}$ (r$_g$)  &   $1.26^{+.34}_{-0.04}$ & Inner disk radius  \\
R$_{out}$ (r$_g$) (fixed) &   $1000$ & Outer disk radius  \\
$\Gamma$ (tied to primary continuum) &   $2.39$   &  Blurred reflection photon index \\
log($\xi$)  [erg cm s$^{-1}$]  & $3.06^{+.03}_{-0.07}$   & Ionization parameter of the disk  \\
Ab$_{Fe}$  &  $1.3^{+.5}_{-0.1}$   & Iron abundance (solar)  \\
log(n$_e$)  [cm$^{-3}$]  & $15.2^{+.2}_{-0.1}$   & Electron density of the disk \\
N$_{blur}$  ($10^{-4}$) & $2.4^{+0.3}_{-0.2}$ &  Normalisation of the relativistic reflector \\ 
\hline
\multicolumn{3}{|c|}{Distant Reflection (XILLVERD)} \\
\hline
$\Gamma$ (tied to primary continuum) &   $2.39$   &  Distant reflection photon index \\
Ab$_{Fe}$  &  $0.61^{+.12}_{-0.07}$   & Iron abundance (solar)  \\
log(n$_e$)  [cm$^{-3}$] (fixed)  & $15.0$   & Electron density of the distant reflector \\
log($\xi$)  [erg cm s$^{-1}$]  (fixed) & 0   & Ionization parameter of the distant reflector  \\
$\theta^o$ (fixed) &   60 & Inclination angle of the distant reflector  \\
N$_{dist}$  ($10^{-4}$) & $0.77^{+0.08}_{-0.07}$ &  Normalisation of the distant reflector \\ 
\hline
\multicolumn{3}{|c|}{Wind Parameters UFO 1 } \\
\hline
log(U) & $2.4\pm 0.1$ &  Ionization parameter  \\
log(N$_H$) [cm$^{-2}$] & $23.3^{+0.1}_{-0.2}$ &  Columns Density  \\
V$_{out}$ (fixed) (km s$^{-1}$)& 44790 & Outflow Velocity \\
$\Delta$ $\chi^2$/$\Delta$$\nu$ &  41.9/2    &  Improvement including UFO 1   \\
\hline
\multicolumn{3}{|c|}{Wind Parameters UFO 3 } \\
\hline
log(U) & $-2.9\pm 0.1$ &  Ionization parameter  \\
log(N$_H$) [cm$^{-2}$] & $20.2^{+0.1}_{-0.2}$ &  Columns Density  \\
V$_{out}$ (fixed) (km s$^{-1}$)& 23610 & Outflow Velocity \\
$\Delta$ $\chi^2$/$\Delta$$\nu$ &  19.1/2    &  Improvement including UFO 3   \\
\hline
\multicolumn{3}{|c|}{Final Statistic } \\
\hline
$\chi^2$/$\nu$ &  998.5/943     & Final statistic \\
\hline
\end{tabular}
\label{tab:epic}
\end{table*}


We then added the different UFO components found in the RGS analysis to this baseline model to test their presence. We left free to vary the ionization parameter and column density of each PHASE component. However, we fixed the outflow velocity to the best fit value obtained in our final model of the RGS data, which will be presented in \S \ref{sec:rgs_ref} (our final description of the RGS data consists of the same, physically motivated continuum model presented here, including relativistic and distant reflection, plus the UFO components). We note that in these models, the distant reflection component (XILLVERD) was not attenuated by the ionized wind, as it should lie further out. 

We find that UFO2 and UFO4 are not required by the EPIC-pn data as adding them to the model results in only a marginal decrease in  $\chi^2$.  However, UFO1 and UFO3 are clearly required  in the fit. The improvement in the statistics after including UFO1 is $\Delta\chi^2  =41.9$ for a difference of 2 d.o.f. The improvement by UFO3 is $\Delta\chi^2  =19.1$/2 d.o.f. 

Table \ref{tab:epic} shows the parameters of our best-fit model, that is presented in Figure~\ref{fig:delchi}. The  UFO parameters are fully consistent with those found over the phenomenological model of the RGS data (see Table \ref{tab:wa_par}), with the exception of the column density of UFO1, who is $\sim$2 times higher in the CCD model. This result will be further discussed in \S \ref{sec:rgs_ref}. The presence of UFO1 is required due to its high opacity produced by bound-bound transitions. On the other hand, UFO3 is required by the fit because of its bound-free opacity above 1 keV. Finally, we note that the continuum parameters are similar those obtained by \citet[][]{Mallick2018}.

The EPIC-pn data do not show evidence of UFO absorption by higher ionization species such as \fexxv\ or \fexxvi. There are no strong residuals between 7.6 and 8 keV, where absorption features by \fexxv\ or \fexxvi\ are expected for gas outflowing at the velocity of UFO1 (see Fig.~\ref{fig:delchi}). This is consistent with the ionization state of UFO1, that predicts no absorption by these charge states. We do not find strong evidence of absorption at other velocities either. 


\section{Modelling the RGS high resolution data with a physically motivated model  \label{sec:rgs_ref}}

The continuum in the high RGS data of Mrk~1044 was well described with a phenomenological model with a powerlaw (\S \ref{sec:powerlaw}). However, the analysis over the EPIC-pn data showed far more complexity than this simple model, including relativistic reflection by the disk.
In this section, we model the RGS data of Mrk~1044 including a physically motivated model for both the UFO and the continuum.

We start by applying the same continuum model applied to the low resolution data. The model consists of a direct emission source (powerlaw) plus relativistic (RELXILLD) and distant reflection (XILLVERD), attenuated by Galactic absorption. We note, however, that the RGS covers only a fraction of the overall spectral range of the EPIC-pn. Then, some of the parameters of the model cannot be well constrained with the RGS alone. To circumvent this problem, we fixed as many parameters as possible in the RGS model to their best fit values on the low resolution data. We find that the RGS can be well fit by a model with only seven free parameters: The spectral index and normalization of the powerlaw, the emissivity index for the outer and inner portions of the disk, as well as the disk break radius and normalization for the RELXILLD component, and finally the normalization for the XILLVERD component. The spectral index of both reflection components was fixed to the same value of the powerlaw component. The rest of the parameters were constrained to the values reported in Table \ref{tab:epic} (a model with these parameters free does not give physically meaningful results, and does not improve significantly the fit). This model gives a $C$ statistic of 2475 for 2157 d.o.f. 

\begin{table*}
\caption{Best fit parameters of the physically motivated continuum and the multiphase ultra-fast wind detected in the RGS spectrum of Mrk~1044.}
\centering
\begin{tabular}{c c c c c c c c}
\hline
\multicolumn{8}{|c|}{Primary Continuum Parameters} \\
\hline
$\Gamma$  &  \multicolumn{2}{c}{Norm. ($10^{-3}$ ph keV cm$^{-2}$ s$^{-1}$)}  & & & AIC & & \\
$2.57\pm 0.07$ & $6.0\pm 0.6$ &  & & & 2489  & &\\
\hline
\multicolumn{8}{|c|}{Relativistic Reflection Parameters$^*$} \\
\hline
$\Gamma$  & Norm. ($10^{-4}$) & $q_{in}$$^{**}$ & $q_{out}$$^{**}$ & R$_{br}$$^{**}$ (r$_g$)  & & & \\
$2.57$(fixed) & $1.1\pm 0.2$ & $7.36\pm 0.45$ & $2.87_{-0.44}^{+0.23}$ & $12.9^{+1.5}_{-4.4}$ & & \\

\hline
\multicolumn{8}{|c|}{Distant Reflection Parameters$^*$} \\
\hline
$\Gamma$  & Norm. ($10^{-4}$) & & & & & &  \\
$2.57$(fixed) & $0.06\pm 0.04$ & & & & & \\
\\
\hline
\multicolumn{8}{|c|}{Wind Parameters} \\
\hline
 Wind Phase  & $\log U$       &  $\log N_{\rm H}$     & $v_{\rm out}$   &   Statistics & AIC & Factor &  Significance  \\
    -  & 
    &  (cm$^{-2}$) &        (km s$^{-1}$) &  $\Delta C$/$\Delta\nu$   &  \\
    \hline
UFO1  &  $2.3\pm0.1$   &  $23.2_{-0.1}^{+0.2}$ & $44790_{-60}^{+140}$ &  37/3  & 2458 & 5$\times10^{6}$ & $8\sigma$  \\
           
UFO2   &  $1.8\pm0.1$  &  $21.2_{-0.12}^{+0.19}$ & $25960_{-230}^{+140}$ &  19/3 & 2445 & 665 &  $4\sigma$ \\ 

UFO3   &   $-3.1\pm0.2$ &  $20.3\pm0.1$  &  $23610_{-250}^{+210}$     &  20/3 & 2431 & 1096 &  $4\sigma$  \\ 

UFO4    &  $-0.1 \pm0.1$ & $20.2\pm0.1$ &  $23420_{-120}^{+190}$   &  16/3 & 2421 & 148 & $3\sigma$   \\ 
\\
    \hline
\multicolumn{8}{|c|}{Final Statistic} \\
    \hline
 &  &  C$_{stat}$/$\nu$ &  2383/2145 & & \\
\hline\hline
\end{tabular}
\tablecomments{$^*$The parameters not quoted in this table were fixed to the values of the EPIC-pn model presented in Table \ref{tab:epic}.\ $^{**}$ The description of these parameters is presented in Table \ref{tab:epic}.}
\label{tab:wa_par_rexill}
\end{table*}

We then proceed to include in the modelling the four UFO components in the same order that were found in \S \ref{sec:results_rgs}. As before, we quantitatively evaluated the improvement of each component to the fit with the aid of the Akaike Information Criterion values, and with Monte Carlo simulations as described in \S \ref{sec:significance}.  We found that the four UFO components are required to fit the data. UFO1 is required at an 8$\sigma$ level with the Monte Carlo methods. The AIC statistics indicate that the model including this component is 5$\times10^6$ times better than the model without it. Although these numbers are smaller than those found with the phenomenological model, this UFO has a very high significance. Monte Carlo simulations show that UFO2, UFO3 and UFO4 are required with significance 4$\sigma$, 4$\sigma$, and 3$\sigma$, respectively. AIC statistics show that the model with UFO2 is better by a factor $\approx$650, the one with UFO3 is better by a factor  $\approx$1100 and the model including UFO4 by a factor $\approx$150. 

Table \ref{tab:wa_par_rexill} presents the results of our best fit. The continuum parameters are similar, yet not equal, than those found in the EPIC-pn data. This is likely because of the much larger sensibility of the high-resolution data to detect individual narrow absorption lines, which results in this model requiring the presence of UFO2 and UFO4. 

It is noteworthy that the best fit values obtained for the wind components are strikingly similar to those obtained with the phenomenological model of the continuum presented in \S \ref{sec:results_rgs}. This shows that the UFO properties are basically defined by the presence of the absorption lines. The only exception to this is the column density of UFO1. As was the case for the EPIC-pn data, the value of the column density was twice that found with a simple powerlaw. We note that this physically motivated continuum model requires a slightly higher continuum in the 6-12\AA\ band (both over the high and low resolution data), where most of the bound-bound opacity of this UFO is present. Thus, the larger column density in this component is required to balance the larger continuum, and fit well the depth of the absorption lines. Nevertheless, the final fit cannot be easily differentiated by eye to the one presented in Figure \ref{fig:rgs4}.

In the following, we discuss our results with the model presented in this section, that includes a physically reasonable description of the continuum and absorption components.

\section{Discussion \label{sec:disc}}

\subsection{Comparison with Previous Studies}
The present analysis confirms the results of \citet[][]{Mallick2018} indicating that the broad-band spectra of Mrk~1044 (including the soft excess) can be well explained by a model consisting of relativistic blurred reflection of the primary continuum source by a high-density accretion disk that has a broken powerlaw emissivity profile. 

We note that \citet[][]{Mallick2018} also report the presence of a UFO, with column density of 4.7$\times10^{20}$ cm$^{-2}$, an outflow velocity of 0.1c, and high ionization state. However, the UFO reported by these authors is remarkably different than the UFO components presented here in that it is detected as a very broad absorption profile ($\sigma\sim$15,000 km $^{-1}$), associated with an \oviii\ feature near 0.7 keV. This absorption feature is found in the EPIC-pn and RGS data after extrapolating the continuum model in the 3-10 keV region to the whole band.

Our models of the data do not require this broad feature (in neither the low and high resolution data). We find that our continuum modelling for EPIC-pn predicts only a modest broad absorption feature near 0.7 keV (see middle panel of Fig. \ref{fig:delchi}). Our final fit, including two UFOs, can account for the most part of the spectral broad-band shape, including the absorption near this energy region (as seen in the lower panel of Fig. \ref{fig:delchi}). Figures \ref{fig:rgs1} and \ref{fig:rgs4} show the same for the RGS. 

We have also found that the depth and width of this broad line-like structure depends on the exact continuum parameters of the relativistic reflection component, and that this feature is not required with a different description of the whole band continuum  (i.e., a model consisting of a powerlaw with a distant neutral reflector plus a ``traditional" soft excess modelled by two blackbodies or by an additional powerlaw). 

The evidence for the reality of the UFO components reported here is based upon the simultaneous detection of several statistically significant narrow absorption lines. These absorption lines can be fit simultaneously with physically motivated self-consistent absorption models, whose properties are basically independent of the continuum modelling.

With this in mind, we consider the UFO description presented here a more reliable representation of the data. We note that the main purpose of the analysis presented by \citet[][]{Mallick2018} was to understand the physical processes behind the continuum emission, while our analysis is centered in understanding the absorbing systems.

\subsection{Ultrafast Outflows and Narrow Line Seyfert 1 Nuclei}
Mrk 1044 represents another case of an NLS1 in which a UFO has been detected. Over the years, there has been an increasing number of  subrelativistic winds in this type of AGN. Other detections include
Akn 564 \citep{Gupta2013},
IRAS\,17020+4544 (L15),
PG\,1211+143 \citep{Reeves2018}, 
1H\,0707-495 \citep{Kosec2018},
1\,Zw\,1 \citep{Reeves2019}, and
IRAS\,13224$-$3809 \citep{Parker2017}. This might not result entirely surprising,  since one of the prevailing ideas for the launching mechanism of these winds is that they are radiatively driven \citep{Matzeu2017} and NLS1 galaxies have suitably high Eddington
ratios. We note, however, that recent results cast doubts on the efficiency of radiation to accelerate gas to subrelativistic speeds \citep[up to 0.1c, ][]{Luminari2021}. In any case, the large fraction of UFOs detected in NLS1 points to an intimate dependence of the UFO launching mechanism, whatever it might be, with the accretion rate.  

\subsection{Multi-Phase UFOs in AGN}
Mrk 1044 shows a complex ultra-fast outflow wind structure consisting of four different absorbing components: a ``heavy'' and very hot absorber with roughly twice the velocity of the other three slower and much lighter components (the ``light'' components). There is a vast range of column densities, differing by up to a factor of $\sim1000$, and ionization states, differing in $U$ by nearly five orders of magnitude, between the hottest and coolest components in the overall outflow. It is noteworthy that, within the errors, two of the light absorbers (UFO3 and UFO4) have the same velocity and the same column density, within a factor of two. This is suggestive of a single system spanning a vast range of temperatures, or a multi-phase outflow.

This overall structure is not unprecedented.  L15 reported a multi-phase UFO  in IRAS\,17020+4544, with a single component dominating the mass and dynamics, plus four lighter components. Strong differences in ionization state and column density among the components were also found in that case. L15 also found  two different UFOs with the same outflow velocity. In comparison, the wind in IRAS\,17020+4544 has half the velocity of that found in Mrk 1044. In that case, all the components have similar velocities, including the ``heavy'' component. In both objects, the lines are narrow and unresolved, pointing to negligible acceleration along the line of sight.  \cite{Gupta2013,Gupta2015} also found evidence of multi-phase UFOs in Ark 564 and Mrk 590.  
Recently,  \cite{Parker2020} observed a multi-phase outflow in IRAS 13349+2438,  with two highly ionized components absorbing in the 
Fe K band, and a lower ionization Ly$\alpha$ line accompanying the slower component. A few other examples of  multiple-ionization UFOs exist, including PDS 456 \citep{Reeves2018} and 
MCG-03-58-007 \citep{Braito2018, Matzeu2019}.

The similar characteristics of the UFOs in these objects, 
particularly Mrk 1044 and IRAS\,17020+4544, introduce a new view of UFOs. They show much more complex structure than a simple wind produced by one or two lines of highly ionized Fe. These results also show that lower ionization UFOs exist, contrary to the expectations of models where the wind is bound to exist only at higher ionization levels \citep[e.g.,][]{King2010}. The  CCD and grating data on Mrk 1044 exclude the presence of such highly ionized material (no \fexxv\ or \fexxvi\ absorption is detected), and show that the nature of UFO1 is the same as that of the UFOs detected in the Fe K band of other objects with similar column densities and velocities,  but with significantly lower ionization. The ``heavy component'' of L15 (UFO C in their paper) may represent an even more extreme case of this.

\subsection{The Nature of the Different Wind Phases}

The multi-phase structure of these outflows offers valuable clues to understanding their nature. A simple scenario where dense blobs of lower ionization material are embedded in a larger scale wind (the ``heavy'' absorber)  is ruled out by the UFOs detected in Mrk 1044, as the difference in velocity among the components, nearly $0.1c$, would produce shear forces that would destroy the blobs on very short timescales. Yet, in some cases, these components persist over timescales of years, e.g., component A in IRAS\,17020+4544 (L15).

In the few cases with multi-phase outflows detected in the Fe K band --- IRAS 13349+2438 \citep{Parker2020}, PDS 456 \citep{Reeves2020}, and  MCG-03-58-007 \citep{Matzeu2019} --- the presence of different components can be explained by a disk wind with different streamlines crossing our line of sight to the central engine, with the more ionized and faster material coming from smaller radii
\citep[e.g.,][]{Fukumura2015}. In these sources, the different velocity components detected in absorption all have high column densities ($N_{\rm H} \approx 10^{23-24}$\,\pcm), and the column densities of the lower ionization material can vary by two orders of magnitude \citep{Reeves2020}.

The striking resemblance of the UFOs in Mrk 1044 and IRAS\,17020+4544, and their difference with respect to the Fe K band multi-phase UFOs, suggests that in this case the mechanism might be different. Given the low ionization states and column densities of the ``light'' components in these objects, different streamlines crossing our line of sight are unlikely. Such light and warm/cold components  arising directly from the accretion disk are difficult to explain, and their  subrelativistic velocities would make them highly unstable. 

The properties of the phases in these two objects might instead be more easily understood in terms of a ``shocked outflow'' scenario. This model posits an initial fast outflow, launched at accretion disk scales with outflow velocity $\sim0.1c$ that shocks the ambient medium \citep[see][]{King2015}. In such a scenario, fluid instabilities (e.g., Rayleigh--Taylor) may form at the interface of the shock, as occurs in supernova remnants
\citep{Velazquez1998} due to the difference between the densities of the impacting wind and the impacted medium. A condition for the Rayleigh--Taylor instability to grow is that the mass of the  ISM that is pushed by the discontinuity should be higher than the mass of the ejecta \citep{Velazquez1998}, which undergoes a deceleration process that is able to trigger instabilities in the fluid. A hydrodynamic toy model simulating the formation of such instabilities within the central region of a NLS1 is presented by \cite{Longinotti2020}, including typical UFO parameters.  The results show that ``plumes'' or ``fingers'' of gas with velocities of the same order of magnitude as the incident UFO are formed as a result of instability processes in the expansion of the shocked outflow. When these ``fingers''  intercept our line of sight to the center,  they can produce  the ``light'' UFO components observed in the spectra \citep[see also][]{Sanfrutos2018}. Furthermore, entrained colder material would naturally explain the presence of the very low ionization states found in UFO3, and its common speed with UFO4. Thus the overall structure of these winds might be explained by this scenario.

 \subsection{Implications for Feedback}
 \label{section:implications}
It is now well established, at least using simple energetic arguments, that UFOs carry enough power to quench star formation in their hosts and influence the evolution of galaxies. According to \cite{Hopkins2010}, a wind transferring energy outwards in an energy-driven mode  requires a minimum power to produce feedback $\sim0.5$\% of the Eddington luminosity.\footnote{ In an ``energy-driven" wind scenario, most of the energy of the UFO is transferred outwards and injected into the insterstellar medium, rather than radiated away}
As we show below, the UFO in Mrk 1044 is not an exception, and  may satisfy this condition. 

We first address the question of the black hole mass. \cite{Du2019} give a
mass of $M = 2.85 \times 10^6\,\msun$ based on reverberation measurements
\citep{Hu2015}. Unfortunately, the only line-width measurement quoted
is FWHM(\hb), which introduces a bias by stretching the mass scale, i.e., low masses are underestimated and high masses are overestimated \citep{DallaBonta12020}. However, it is possible to remove this bias empirically. We take from \cite{Du2019}
the luminosity of the \hb\ line and adjust it to the cosmological parameters used by \cite{DallaBonta12020}, i.e., $H_0 = 72$\,\kms\,Mpc$^{-1}$, $\Omega_\Lambda =0.7$, $\Omega_{\rm m} = 0.3$, which gives $\log L(\hb) = 41.33 \pm 0.09$, where $L(\hb)$ is in ergs per second. Taking the line width ${\rm FWHM} = 1178 \pm 22$\,\kms from \cite{Du2019} and using equations (40) and (41) of \cite{DallaBonta12020} gives
$M = 3.78\,(\pm 1.45) \times 10^6\,\msun$. The corresponding Eddington luminosity is $L_{\rm Edd} = 1.26 \times 10^{38} (M/\msun) = 4.76 \times 10^{44}$\,\ergsec.

To estimate  the energetics of the outflow in Mrk 1044, we assume that the outflow velocity is larger than or equal to the escape velocity at the launch radius. We take this to be  $r = 2G M/v_{\rm out}^2$ Under this condition, the mass outflow rate for a radial wind in spherical geometry crossing our line of sight can be estimated by
\begin{equation}
\dot{M}_{\rm out} =4\pi \mu m_p r N_{\rm H} v _{\rm out}  C_f ,
\end{equation}
where $m_p$ and $\mu$ correspond to the mass of the proton and to the mean atomic mass per particle ($\mu=1.4$), $C_f$ to the covering fraction of the  outflow, and $N_{\rm H}$ and $v_{\rm out}$ to the column density and the observed velocity of the outflow, respectively \citep{Tombesi2015}.
Following L15, we can express this formula in units of 10$^{21}$~cm$^{-2}$ 
($N_{\rm H} = N_{21} \times 10^{21}\,{\rm cm}^{-2}$)
and $10,000\,\kms$
($v_{\rm out} = v_{10} \times 10^4\,\kms$)
as 
\begin{eqnarray}
\label{eq:outflow}
\dot{M}_{\rm out} & = & 7.80\times 10^{15} \left( \frac{M}{M_\odot}\right) 
\left(\frac{N_{21}}{v_{10}}\right) C_f\,{\rm g\,s}^{-1} \nonumber \\
& = & 1.24 \times 10^{-10} \left(\frac{M}{M_\odot}\right) 
\left( \frac{N_{21}}{v_{10}}\right) C_f M_\odot\,{\rm yr}^{-1}.
\end{eqnarray}
The energy outflow rate, or kinetic luminosity, is

\begin{multline*}
\dot{E}  =  \frac{1}{2} \dot{M}_{\rm out} v_{\rm out}^2 \\
=  3.9 \times 10^{33} \left( \frac{M}{\msun} \right)\ N_{21}\ v_{10}\ 
C_f\ \ergsec.
\end{multline*}

We find that UFO1 is the only component that could produce feedback. Using
$N_{\rm H}$ and $v_{\rm out}$ from Table \ref{tab:wa_par_rexill} yields
\begin{equation}
\dot{M}_{\rm out} \approx 1.6 \times 10^{-2} C_f\ {\rm \msun\,yr}^{-1},
\end{equation} 
and 
\begin{equation}
\dot{E} = 1.0 \times 10^{43}\ C_f \ \ergsec.
\end{equation}
The \cite{Hopkins2010} criterion for triggering feedback by an energy-conserving wind is
\begin{equation}
\frac{\dot{E}}{L_{\rm Edd}} = 0.02 C_f > 0.005,
\end{equation}
which is satisfied for $C_f > 0.25$. We note that this value bares caveats associated with the location of the wind that goes in the estimation of $\dot{E}$, and with the black hole mass which goes in the estimation of $L_{\rm Edd}$, as discussed above. Thus, the above ratio should be taken with care.

The studies of \citet[][]{Tombesi2010} and \citet[][]{Gofford2013} found that the fraction of Seyfert galaxies with UFOs is about 30-40\%. If this fraction is interpreted as a proxy for the global covering factor of the wind, the UFO in Mrk~1044 would meet the theoretical feedback condition for a wind in the energy-conserving mode. Although these are estimates based on strong assumptions, an emerging picture seems to be arising where winds in NLSy1 might be capable of strong feedback.

On the other hand, if most of the thermal energy released at the shock front of the UFO with the interstellar medium is radiated away, the wind in Mrk~1044 would not have a major transforming effect on its host galaxy, as the criterion for feedback in the ``momentum-conserving" regime is much more demanding: ${\dot{E}}/{L_{\rm Edd}}~>~0.1$.

We note, however, that even if the wind falls short to produce strong feedback and completely halt star formation, it could still have important effects on the host galaxy. For instance, if the wind could expand above and below the disk (directions of least resistance), it could inflate  bubbles in the circumgalactic medium of the galaxy on timescales of millions of years, similar to the so-called Fermi bubbles in the Milky Way
(\citealt{Faucher2012}; see \citealt{Nicastro2016} for further details). However, we note that if this were the case, the feedback is only in its very early stages, as no evidence for such bubbles is found in the analysis of MUSE optical data of Mrk~1044 \citep{Powell2018}. An early stage in the feedback process is also suggested by the lack of X-ray warm absorbers in Mrk~1044, because the ``outflow shock model'' predicts the formation of these systems as the shocked gas cools down and slows down. 

The UFOs in  IRAS\,17020+4544 and Mrk 1044 show that feedback is not exclusive to the most luminous active galaxies in the Universe 
\citep[e.g.,][]{Chartas2002,Tombesi2015,Nardini2015},
but can take place even in Seyferts with moderate luminosity.

\section{Conclusions}
Mrk 1044 provides evidence for a bona-fide ultra-fast outflow in the soft X-ray spectrum of an NLS1 galaxy. The increasing number of UFOs detected in NLS1s suggests an intimate relation between the high accretion rates in this type of AGN, and the mechanism responisble for the launching of the winds.  The nature of the outflow in Mrk 1044, with a multi-velocity, multi-phase structure, is strikingly similar to that found in IRAS 17020+4544. It appears that the structure of subrelativistic outflows is more complex than thought before. The presence of a ``heavy'' hot component supports the idea of a ``shocked outflow''  where the lighter and warmer absorbing components form by instabilities in the shock. The energetics of the wind could be sufficient to produce strong feedback effects on the host galaxy in the so-called ``energy-conserving mode", showing (along with  IRAS 17020+4544) that such feedback  might not be exclusive of the most luminous active galaxies in the Universe, but can take place even in Seyfert galaxies of moderate luminosity.

\acknowledgments
We thank the anonymous referee for his/her useful comments. YK acknowledges support from DGAPA-UNAM grant IN106518, DGAPA-PASPA 2016-2017 and from the Faculty of the European Space Astronomy Centre (ESAC).  ALL acknowledges support from CONACyT grant CB-2016-286316.

\vspace{5mm}
\facilities{XMM-Newton}

\software{SAS (v18.0.0; Gabriel et al. 2004), XSPEC (v12.11; Arnaud 1996), PHASE (Krongold et al. 2003)}







\bibliography{ms}{}
\bibliographystyle{aasjournal}



\end{document}